# Nature as the observer: A simplified approach to the measurement problem and related issues

Fred H. Thaheld*

## Abstract

Several theories have been advanced recently which appear to offer a resolution to that portion of the measurement problem which previously dealt with a possible reduction of the state vector in a subjective fashion by the brain, mind or consciousness. It now appears, based on both biological and mathematical analysis, that collapse of the wave function always takes place in an *objective* fashion in the retinal rod-rhodopsin molecule, and that only measured information is ever presented to the brain, mind or consciousness for possible *subjective* analysis. The remaining portion of the measurement problem has to do with the use and legitimacy of such terms as *boundary (Heisenberg* or *von Neumann cut), information, irreversible, measurement, microscopic, macroscopic, observer (apparatus* or *measuring system), observed (measured system)* and *wave function collapse.* This portion of the measurement problem may be resolved in two ways. First, by adopting Dirac's theory that it is *nature* that makes the choice of measuremental result. Second by the insertion of a *mesoscopic* bridge between the *microscopic* and *macroscopic* worlds, in the existing form of the rhodopsin molecule with its retinal and opsin components.

*fthaheld@directcon.net

Introduction

First, let me say once again that, as per the very nice paper of Bell (1), I agree that there are more or less 'measurement-like' processes going on more or less all the time, more or less everywhere without human intervention. The only thing I would add to this is that these processes can involve either *animate* or *inanimate* entities, with emphasis on the *animate* in this paper. Second, that while he objected to certain words which he felt have no place in a *formulation* of some serious part of quantum mechanics with any pretension to physical precision, he comes down the hardest against the term 'measurement'.

In previous papers I have attempted to deal with part of this contentious issue of 'measurement' by demonstrating that neither the brain, mind or consciousness has any *subjective* role to play in the collapse of the wave function and, that this takes place in an *objective* fashion in the retinal rods, with emphasis upon the rhodopsin molecule (2-4). By way of explanation, rhodopsin is the molecule that takes part in the initial step in the vision process and has two components, 11-*cis* retinal (which can absorb a photon) and opsin, which is a protein molecule (2,3). It is of interest to note here that opsin, or molecules that detect light, appeared in biological systems long before eyes.

An earlier theory advanced by Shimony (5) states that the locus of reduction is the macromolecules of the photoreceptor protein of the rod cells of the eye, rhodopsin. In the resting state of retinal, hydrogen atoms attached to the eleventh and twelfth carbon atoms, lie on the same side of the carbon backbone (so that the conformation is called *cis*), and this arrangement causes the backbone to bend. There is a potential barrier between the *cis* and *trans* conformations, in which the two hydrogen atoms mentioned are on opposite sides of the backbone from each other. But, when retinal in the *cis* conformation absorbs a photon, it acquires sufficient energy for a rotation to occur between the eleventh and twelfth carbon atoms, so that the *trans* conformation is achieved. Shimony's conjecture is that the reduction occurs at the retinal molecule itself; that there is a superselection rule operative which prevents a superposition of molecular conformations as different as *cis* and *trans* from occuring in nature.

Matsuno (2) has commented upon the isomerization of rhodopsin from the standpoint of an internal measurement: "Robust transformation of a quantum such as a *cis-trans* transformation is an example of measurement internal to a molecule. Measurement internal to a molecule is an activity of breaking a spacetime continuum on the part of the interacting electrons and atoms. The atoms provide the potential or the spacetime curvature to the moving electrons, and the electrons then exert the forces of push or pull upon the atoms. These two are not synchronous but sequential. If they are taken to be synchronous, the whole dynamic scheme develops in a unitary fashion as obeying the rule of linear quantum mechanics. In contrast, when these two movements of atoms and electrons are legitimately taken to be sequential, the atomic displacement following the electronic displacement, such as the destablizing steric forces following the electronic excitation of a molecule (rhodopsin), is an indication of measurement proceeding internally. A consequence of such internal measurement is the appearance of a new discontinuity in the former spacetime continuum, like a *cis-trans* transformation".

A modified CSL amplification theory has recently been advanced by Adler (4,6), which also involves the retinal rods, moving the collapse process out of the nervous system or visual cortex. Evidence is thus beginning to accumulate that only *objective* information is ever presented to the brain, mind or consciousness for possible *subjective* analysis.  One can now state the following as a special law of nature:

*No superposed photon state can ever get past the retinal rods of any eye.*

When I say "any eye", this pertains to the eyes of *any* living entity, whether one thinks that the eye evolved just once or repeatedly, since *all* animal eyes share a common molecular strategy using opsin for catching photons (7).

Now that this portion of the measurement problem has been moved from the *subjective* (brain, mind or consciousness) to the *objective* (retinal rod-rhodopsin molecule), the time has come to address the other remaining issues relating to the measurement problem.  This concerns the use of such terms as *boundary (Heisenberg* or *von Neumann cut), information, irreversible, measurement, microscopic, macroscopic, observer (apparatus* or *measuring system), observed (measured system)* and *wave function collapse* (1).  Is it possible to give a tangible meaning to some or all of these terms?

Nature as the 'observer'!

It is of historic interest to note here that Dirac (8), who invented the idea of *wave function collapse*, said that it is *nature* that makes the choice of measuremental result; once made the choice is 'irreversible and will affect the entire future state of the world' (9).  At the same time in 1927, Born agreed with Dirac and said that this was in accordance with the views of von Neumann, whose book on this subject was not yet published (2).  Upon closer examination, one comes to the conclusion that this is a very profound statement from Dirac, with implications for a possible resolution of the measurement problem.  This means that any sentient or non-sentient beings (with emphasis upon humans) are not really *observers* and perform no measurements and, that this process of observation or measurement has been carried out by *nature* for a period of ~ 3.5 billion years in an *objective* fashion as specifically regards any animate entities.

One then arrives at a startling conclusion, that what we have envisioned as being an *observation* or *measurement* is really an *illusion!* And, that all we or any other living entity do is *analyze* or *interpret* and *act upon* the collapsed objective information presented to us by *nature* to the best of our abilities. That the same information is presented to us as it is to the 2 billion year old *Euglena* with its primitive eyespot or, to the ~ 3.5 billion year old photosynthetic *Cyanobacteria.* Once one takes this approach, a whole new vista opens up as regards both quantum mechanics and the classical world (2). Let us explore this in the following fashion:

When a superposed photon state (the *microscopic observed* or *measured system)* impinges upon the *mesoscopic* light-absorbing compound retinal of the rhodopsin molecule (the *observer* or *measuring system)*, this absorption of light by retinal causes a change in the three-dimensional structure of rhodopsin, which is of an *irreversible* nature, since the retinal molecule goes from a "bent configuration" (11–*cis* retinal) to a "straightened configuration" (*trans* retinal). This configuration, most importantly, does not then fit into the binding site of the opsin molecule (2). In addition, while the molecule is changing its shape, a vibrational spectrum is also evolving. It is this combination of the rhodopsin molecular conformational changes accompanied by vibrational motions (vibronic coupling) involving Raman and Franck-Condon considerations, which should lead to a *wave function collapse*.

What this means is that the *microscopic* photon is being absorbed by a contiguous *mesoscopic detector* or the retinal *observer*, proceeding in a more seamless fashion, to a *macroscopic* change or *measurement* in the three-dimensional structure of rhodopsin, which may also involve a transduction and amplification process in the retinal rod (3-4, 6).

The usual process of *wave function collapse* (in the inanimate world) has been viewed as proceeding from *microscopic* to *macroscopic*, which is much more abrupt than the seamless *microscopic-mesoscopic-macroscopic* process envisioned herein (3). It appears in this specific instance that we are looking at a 2-stage *detector* or *apparatus*, with a *mesoscopic* first stage of retinal and a *macroscopic* $2^{nd}$ stage consisting of the rhodopsin molecule and the retinal rod. Both stages of the *mesoscopic-macroscopic detector* or *apparatus*, are always in a specified initial state, *never* affected by the decohering environment, except as follows.

Rhodopsin molecules are so resistant to dechoerence that there are only two ways that they can be activated. First, by successful absorption of a photon by retinal. Second, with excitation originating from thermal isomerization of rhodopsin, which are occasional and spontaneous discrete events resembling single photon responses that occur once every 90 seconds in a mammalian rod. Since each mammalian rod contains approximately $10^8$ rhodopsin molecules, each rhodopsin molecule activates spontaneously approximately *once every 300 years* (2,3)!

By inserting this *mesoscopic* retinal (which is immune to any environmental decoherence) between the *microscopic* and the *macroscopic*, we may be opening the door to a legitimate empirical examination of the quantum mechanical and classical worlds, thereby lending credence to the concept of a *Heisenberg* or *von Neumann cut* while at the same time making the concept of *wave function collapse* a much more benign affair than was previously thought possible. In fact, if it is now a more gentle process, this may help to remove some of the objections to its existence.

Matsuno (10) has recently commented upon the argument for an objective wave function collapse in the following fashion: "In short, this means the interplay between the Hilbert space geometry and ordinary spacetime geometry. Isomeric transformation of a molecule (rhodopsin) must be a case to the point. Biomolecules such as rhodopsin and microtubule can demonstrate the interplay between the Hilbert space and the ordinary phase space even if we do not enter into the Planck scale. I found your argument is powerful enough".

Conclusion

1. That a measurement is performed by *nature* serving in the role of an *observer* in the retinal rod, more specifically the rhodopsin molecule, leading to a stochastic discontinuous nonlinear *objective wave function collapse.* This means that while neither the brain, mind or consciousness play any *subjective* role in the collapse of the wave function, they do play a role in analyzing and interpreting the information presented to them.

2. Even though the eye (which came before the brain) and, by association, the retinal rod-rhodopsin molecule-retinal, is an integral part of the brain, it is a non-conscious entity, such that the brain, mind or consciousness can have no *subjective* effect upon its *objective* and stochastic measurements.

3. That *wave function collapse* is a real physical process, resulting when a superposed *microscopic system* (photon) interacts with a *mesoscopic-macroscopic apparatus* (rhodopsin molecule), with *irreversibility* occuring at the *mesoscopic* level as a result of the *cis-trans* conformational change. As a result of this *mesoscopic* bridge, the actual collapse process may be much gentler.
4. That one may now be able to subject to empirical test the concept of *boundary* or the *Heisenberg-von Neumann cut* between the quantum and classical worlds, by virtue of this *mesoscopic* bridge connecting these two worlds. And, that this *cut* occurs in the same place measurement after measurement, in the rhodopsin molecules in the rod cells of the retina.
5. If an actual *wave function collapse* occurs, then a superposed state is not the final outcome of a measurement or observation by *nature*, which means that the world state vector does not split into two branches. This renders Everett's theory regarding the existence of many worlds, minds or universes untenable, at least when it is presented in this fashion.
6. If the Schroedinger linear equation has to be modified to include nonlinear discontinuous changes, then some definitions of physical units may also be affected, opening up a whole new domain of empirical and theoretical physics, with implications in the area of quantum gravity (2). The modification could be of a minor nature if the collapse is more gentle. This raises the intriguing possibility that 'collapse' may not now be the proper terminology.

Amazingly, we now end up having lent some credence to some or all of the above terms, which were previously considered to be of an ambiguous, divisive or ephemeral nature. We observe *nothing*, rather *observed* or *collapsed information* of an *objective* nature is presented to us in a sequential fashion by *nature* for our *subjective* analysis or interpretation. Our supposed role as an *observer* is just an illusion!

## References

1. Bell, J.A., 1990. Against 'measurement'. Phys. World. Aug. 33-40.
2. Thaheld, F.H., 2005. Does consciousness really collapse the wave function? A possible objective biophysical resolution of the measurement problem. BioSystems 81, 113-124. quant-ph/0509042.
3. Thaheld, F.H., 2006. The argument for an objective wave function collapse: Why spontaneous localization collapse or no-collapse decoherence cannot solve the measurement problem in a subjective fashion. quant-ph/0604181.
4. Thaheld, F.H., 2006. Are we getting closer to a resolution of the measurement problem? quant-ph/0607127.
5. Shimony, A., 1998. Comments on Leggett's "Macroscopic Realism", in *Quantum Measurement: Beyond Paradox.* R.A. Healey and G. Hellman, eds. Univ. Minnesota, Minneapolis. p.29.
6. Adler, S., 2006. Lower and upper bounds on CSL parameters from latent image formation and IGM heating. quant-ph/0605072.
7. Fernald, R.D., 2004. Evolving eyes. Int. J. Dev. Biol. 48, 701-705.
8. Dirac, P.A.M., 1930. *The Principles of Quantum Mechanics.* Clarendon, Oxford.
9. Kragh, H., 1990. *Dirac: A Scientific Biography.* Cambridge Univ. Press, Cambridge.
10. Matsuno, K., 2006. Personal communication.